\documentclass{emulateapj}
\usepackage{apjfonts}
\newcommand{\junits}{\mbox{ erg cm$^{-2}$ s$^{-1}$ Hz$^{1}$ sr$^{-1}$}}

\newcommand{\eV}{\mbox{ eV}}
\newcommand{\kel}{\mbox{ K}}

\newcommand{\Msun}{\mbox{ M$_\odot$}}
\newcommand{\Zsun}{\mbox{ Z$_\odot$}}

\newcommand{\hunits}{\mbox{ km s$^{-1}$ Mpc$^{-1}$}}
\newcommand{\kms}{\mbox{ km s$^{-1}$}}
\newcommand{\bq}{\begin{equation}}
\newcommand{\eq}{\end{equation}}
\newcommand{\bqa}{\begin{eqnarray}}
\newcommand{\eqa}{\end{eqnarray}}

\newcommand{\bxi}{\bar{x}_i}

\newcommand{\hii}{\ion{H}{2} }

\newcommand{\zthree}{\zeta_{\rm III}}
\newcommand{\ztwo}{\zeta_{\rm II}}

\newcommand{\ptwo}{p_{\rm II}}
\newcommand{\mmin}{m_{\rm min}}
\newcommand{\fnew}{f_{\rm new}}
\newcommand{\fcoll}{f_{\rm coll}}
\newcommand{\deriv}{{\rm d}}

\begin{document}

\title{Is Double Reionization Physically Plausible?}

\author{Steven R. Furlanetto\altaffilmark{1} \& Abraham
  Loeb\altaffilmark{2}} 

\altaffiltext{1} {Division of Physics, Mathematics, \& Astronomy;
  California Institute of Technology; Mail Code 130-33; Pasadena, CA
  91125; sfurlane@tapir.caltech.edu}

\altaffiltext{2} {Harvard-Smithsonian Center for Astrophysics, 60
Garden Street, Cambridge, MA 02138; aloeb@cfa.harvard.edu}

\begin{abstract}

Recent observations of $z \sim 6$ quasars and the cosmic microwave
background imply a complex history to cosmic reionization.  Such a
history requires some form of feedback to extend reionization over a
long time interval, but the nature of the feedback and how rapidly it
operates remain highly uncertain.  Here we focus on one aspect of this
complexity: {\it which physical processes can cause the global ionized
fraction to evolve non-monotonically with cosmic time?}  We consider a
range of mechanisms and conclude that double reionization is much less
likely than a long, but still monotonic, ionization history.  We first
examine how galactic winds affect the transition from metal-free to
normal star formation.  Because the transition is actually spatially
inhomogeneous and temporally extended, this mechanism cannot be
responsible for double reionization given plausible parameters for the
winds.  We next consider photoheating, which causes the cosmological
Jeans mass to increase in ionized regions and hence suppresses galaxy
formation there.  In this case, double reionization requires that
small halos form stars efficiently, that the suppression from
photoheating is strong relative to current expectations, and that
ionizing photons are preferentially produced outside of previously
ionized regions.  Finally, we consider H$_2$ photodissociation, in
which the buildup of a soft ultraviolet background suppresses star
formation in small halos.  This can in principle cause the ionized
fraction to temporarily decrease, but only during the earliest stages
of reionization.  Finally, we briefly consider the effects of some of
these feedback mechanisms on the topology of reionization.

\end{abstract}

\keywords{cosmology: theory --- galaxies: evolution --- intergalactic medium}

\section{Introduction}
\label{intro}

The epoch of reionization marks the time when collapsed objects began
to influence the diffuse intergalactic medium (IGM) and first rendered
it transparent to ultraviolet photons.  It is an important milestone
signaling the end of the ``cosmological dark ages'' and the emergence
of the first luminous sources in the universe.  A wide variety of
observational probes have recently been used to constrain not only the
reionization process itself but also the properties of the sources
driving it.

The most powerful observational constraints come from Ly$\alpha$
absorption spectra of high redshift quasars
\citep{becker,fan,white03}.  At least one $z > 6$ quasar shows a
complete \citet{gp} trough as well as a rapidly evolving neutral
fraction indicative of the final stages of reionization (though see
\citealt{songaila04} for a different interpretation).  This conclusion
is strengthed by analyses of the proximity effects around these
quasars \citep{wyithe04-prox,mesinger04}.  The second clue is from
measurements of the large scale polarization anisotropies of the the
cosmic microwave background (CMB), which imply a high optical depth to
electron scattering and require reionization to begin at $z \ga 14$
\citep{kogut03,spergel03}, albeit with large error bars.  A third clue
comes from the temperature of the Ly$\alpha$ forest at $z \sim 3$,
which is relatively high and indicates that the neutral fraction
changed substantially at $z \la 10$ \citep{theuns02-reion,hui03}.
However, this constraint is difficult to interpret because
helium reionization also heats the IGM (e.g., \citealt{sok02}).

Taken together, these constraints imply a complex and extended
reionization history, ruling out the simplest models in which the
emissivity is a function purely of halo mass and independent of cosmic
time (e.g., \citealt{barkana01} and references therein).  Within the
context of the cold dark matter model, measurements of the matter
power spectrum imply that structure formation occurred extremely
rapidly at high-redshifts: that is, the fraction of baryons
incorporated into star-forming halos at $z=6$ is over three orders of
magnitude larger than the fraction at $z=20$.  If the source
properties remained constant, this exponential growth of structure would
require reionization to be completed within a short time interval.
Thus, if reionization is to \emph{begin} at $z=20$ but \emph{end} at
$z=6$, the sources themselves must evolve significantly during
reionization.

Such evolution could occur naturally through stellar feedback, because
the first galaxies form out of metal-free gas and have exceedingly
shallow gravitational potential wells.  A number of models have been
developed to reconcile the disparate inferences about reionization
\citep{wyithe03,wyithe03-letter,cen03,cen03-letter,
haiman03,sokasian04,fukugita03,somerville03,onken04}, all
incorporating some sort of feedback mechanism to decrease the ionizing
efficiency of the sources.  The resulting ionization histories display
a wide range of features and can extend over long redshift intervals.
One of the more intriguing possibilities is so-called {\it double
reionization}.  We will use this term to designate histories in which
the global ionized fraction $\bxi$ decreases with cosmic time over
some interval.  Such models provide the clearest signatures of
feedback and should be the easiest to observe with future 21 cm
tomography measurements \citep{furl04b}.  They can be contrasted with
``stalling'' models in which $\bxi$ increases monotonically but the
feedback mechanism still shapes the overall evolution.

In this paper, we will critically examine the plausibility of the
physical processes that may lead to double reionization.  Existing
models attribute double reionization to three different feedback
mechanisms.  The first is the transition from metal-free Population
III (hereafter Pop III) star formation to ``normal'' Population II
(hereafter Pop II) star formation as observed in the local universe.
If Pop III stars are massive, they can produce approximately an order
of magnitude more ionizing photons per baryon than do normal stars
\citep{bromm-vms}.  The first stars therefore efficiently ionized the
IGM; however, as Pop III stars died and exploded, they expelled metals
and enriched the IGM.  Once star-forming regions reached a typical
metallicity of $Z \sim 10^{-3.5} \Zsun$, the excess cooling provided
by the metals could reduce the Jeans mass and switch the
star-formation mode from Pop III to Pop II
\citep{bromm01,bromm03-met}, thus lowering the ionizing
efficiency. Under some circumstances, Pop II stars could no longer
counteract recombinations and so $\bxi$ would decrease.
Unfortunately, existing models have treated this transition crudely.
Both \citet{cen03} and \citet{wyithe03} assigned a single, universal
redshift at which star formation switches between the two modes; the
redshift was taken to be the time when the mean metallicity of the
universe passed the above threshold.  Both found that double
reionization occurred for a range of input parameters.  However,
different regions of the universe are expected to reach the same
evolutionary stage of structure formation at different cosmic times
due to modulation by inhomogeneities on large scales; this cosmic
variance is particularly large at high redshifts \citep{barkana04}.
\citet{haiman03} pointed out that if the Pop III/Pop II transition is
spread over a similarly long time interval, double reionization no
longer occurs.  However, they did not calculate the plausible duration
of the transition, and they conflated the metallicity transition with
a change in the mass threshold of galaxy halos for star formation.  In
\S \ref{wind}, we will explicitly compute when and how rapidly the Pop
III/Pop II transition occurs using a physical model for enrichment by
galactic winds.  Like \citet{scann03}, we find that the transition
must occur over an extended redshift interval.  In such a scenario,
double reionization requires a number of unlikely assumptions.

A second feedback mechanism is photoheating.  Reionization raises the
IGM temperature from $T \la 100 \kel$ to $T \ga 10^4 \kel$, increasing
the ambient pressure and hence the cosmological Jeans mass
\citep{rees86,efstathiou92}.  As a result, low-mass halos can no
longer collapse in ionized regions and the star formation rate may
decrease sharply.  The degree of suppression is not clear.  Early
work suggested that this mechanism prevents halos with circular
velocities $V_c \la 30$--$50 \kms$ from forming
\citep{thoul96,kitayama00}.  However, \citet{dijkstra04a} showed
that the suppression is considerably weaker near the time of
reionization, because in that case many halos have already begun to
collapse and their high densities shield them from the ionizing
background.  At the same time, existing halos with virial temperatures
$T_{\rm vir} \la 10^4 \kel$ will photo-evaporate as they absorb
ionizing photons \citep{barkana99, shapiro04}.  The net effect is
that, as reionization proceeds, the mass threshold for galaxy
formation increases by some (uncertain) amount and reionization slows
down.  Again, in such a situation it is possible for the ionizing
sources to ``overshoot'' and for recombinations to dominate for a
time.  This mechanism has been included in most existing models and
rarely been found to produce double reionization.  Nevertheless, there
has been no systematic study of the requirements for such a phase.  In
\S \ref{rad}, we show that photoheating can in principle cause double
reionization, but only in exceptionally optimistic circumstances.  

A third and final feedback mechanism is the photo-dissociation of
H$_2$.  Rotational transitions of H$_2$ provide a cooling channel that
operates in halos with $T_{\rm vir} \ga 200 \kel$
\citep{haiman96,tegmark97}, well below the threshold at which atomic
cooling becomes efficient.  The first halos to form stars in any
hierarchical model have small masses and therefore rely on H$_2$ for
their cooling \citep{abel,bromm02}.  However, H$_2$ is fragile and is
easily dissociated by soft UV photons in the Lyman-Werner band
($11.26$--$13.6 \eV$) \citep{haiman97}.  Once the first stars build up
a sufficient UV background, this cooling channel terminates and the
minimum halo mass to form stars increases; operationally this
mechanism is similar to photoheating although it operates at a lower
halo mass scale.  We will consider whether this scenario can lead to
double reionization in \S \ref{h2}.

Of course, by suppressing the ionizing efficiency in biased regions
these feedback mechanisms affect not only the global reionization
history but also its topology.  As recently stressed by \citet[
hereafter FZH04]{furl04a}, the topology of reionization is observable
through 21 cm tomography (see also \citealt{furl04b}) and through
Ly$\alpha$ absorption \citep{furl04c,wyithe04-bub}.  Although a much
more difficult problem to approach analytically than the global
evolution, it is also worth considering what signatures one might
expect to see in the bubble size distribution from these feedback
mechanisms.  We briefly consider this question in \S \ref{bubbles} and
then discuss other consequences of our results in \S \ref{disc}.

Throughout our discussion we assume a cosmology with $\Omega_m=0.3$,
$\Omega_\Lambda=0.7$, $\Omega_b=0.046$, $H_0=100 h \hunits$ (with
$h=0.7$), $n=1$, and $\sigma_8=0.9$, consistent with the most recent
measurements \citep{spergel03}.  We will use comoving units unless
otherwise specified.

\section{A Metallicity Transition}
\label{wind}

As noted above, existing models attributing double reionization to the
evolving cosmic metallicity rely on an instantaneous transition
between the two modes.  In reality, there are two mechanisms that will
slow it down: the formation of stars in pre-enriched halos (even early
on) and the finite velocity of winds.  Here we will examine these two
processes in turn and apply them to the reionization history.

\subsection{The Formation Rate of New Halos}
\label{fnew}

We divide newly-collapsed gas (i.e., baryons that have just been
incorporated into galaxies able to form stars) into two components.
First, some fraction of the gas accretes onto existing halos that have
already formed stars.  Provided that an existing halo had already
enriched itself beyond the threshold metallicity $Z_t$ to form Pop II
stars, this component of newly collapsing gas cannot make Pop III
stars.  The metallicity threshold is $Z_t \sim 10^{-3.5} \Zsun$
\citep{bromm01,bromm03-met}, which requires $f_\star \ga 10^{-5}
(0.5/f_Z) (Z_t/10^{-3.5} \Zsun)$, where $f_\star$ is the star formation
efficiency and $f_Z$ is the fraction of the stellar mass converted
into heavy elements.  \citet{heger02} find $f_Z \sim 0.5$ for stars
which undergo pair-instability supernovae, although not all massive
stars actually explode so this is only an upper limit to the true
value.  The fraction is somewhat smaller (but still $\ga 0.1$) for
normal initial mass functions (IMFs).  The metallicity threshold for
Pop II star formation is sufficiently small that we expect \emph{any}
halo that has previously form stars to make only Pop II stars after
the initial starburst, unless the metals remain highly clumped in the
halo.  Because clumping would only slow down the transition between
the two star formation modes, we will conservatively ignore it.

To form Pop III stars, gas must therefore collapse into a ``new'' or
``fresh'' halo that is forming stars for the first time.  We take this
criterion to be a minimum galaxy mass $\mmin$ determined by the
physics of radiative cooling.  We will typically assume $\mmin$
corresponds to $T_{\rm vir} = 10^4 \kel$, the threshold temperature
above which atomic hydrogen cooling is effective (e.g.,
\citealt{barkana01}).  Thus, to find the \emph{maximal} rate at which
Pop III stars can form, we wish to know the rate at which halos of
mass $m \approx \mmin$ form.

Unfortunately, as described in the Appendix, computing this rate is beyond the capability of existing semi-analytic models.  We therefore make a simple estimate \citep{sasaki94,verde01}. Given the \citet{press} mass function, we can compute the total rate
of change of the halo mass function:
\bq
\frac{\deriv n(m)}{\deriv z} = n(m) \ \left| \frac{\deriv
  \nu}{\deriv z} \right| \,
\left( \nu - \frac{1}{\nu} \right),
\label{eq:dndz}
\eq 
where $\nu = \delta_c(z)/\sigma(m)$, $\delta_c(z)$ is the critical density for collapse, and $\sigma^2(m)$ is the variance of the density field smoothed on the mass scale $m$.  This total rate of change is the difference between the creation and destruction rates of halos.  The two terms in the derivative can be conveniently (if only approximately) identified as these two rates.  For rare, massive halos (with $\nu \gg 1$), the creation term should dominate because there are only a small number of larger objects into which the halo can be absorbed.  The creation rate should therefore increase with increasing $\nu$.  Halos with $\nu \ll 1$, on the other hand, are already common.  As the nonlinear mass scale rises, these small halos must accrete onto more massive objects. Thus we expect the destruction term to increase with decreasing $\nu$.  The two terms in equation (\ref{eq:dndz}) have the correct limiting behavior, although of course the split does not capture all of the physics, particularly when $\nu \sim 1$.  The rate at which genuinely new halos form is approximately
\bq 
\frac{\deriv \fnew}{\deriv z} \approx \left(
\frac{\mmin^2}{\bar{\rho}} \right) \ \nu \ n(m) \ \left| \frac{\deriv
\nu}{\deriv z} \right|,
\label{eq:dfnew}
\eq
where $\bar{\rho}$ is the mean cosmic density.

\begin{figure}
\plotone{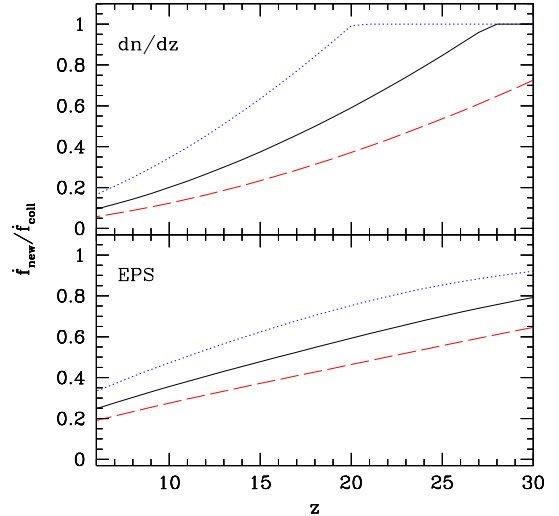}
\caption{Evolution of the fraction of gas collapsing into new halos.
\emph{Top:} Approximate model based on eq.~(\ref{eq:dfnew}).
\emph{Bottom:} Extended Press-Schechter model (see Appendix).  In each
panel, the solid lines have our default $\mmin$ (corresponding to
$T_{\rm vir}=10^4 \kel$).  The dashed and dotted curves correspond to
decreasing or increasing $\mmin$ by an order of magnitude,
respectively.}
\label{fig:fdot}
\end{figure}

The top panel of Figure~\ref{fig:fdot} shows the evolution of $\deriv \fnew/\deriv z$ relative to the total rate of collapse $\deriv \fcoll/\deriv z$, where
\bq
\fcoll = {\rm erfc} \left[ \frac{\delta_c(z)}{\sqrt{2} \sigma_{\rm
      min}} \right]
\label{eq:fcoll}
\eq 
is the fraction of matter bound to halos with $m>\mmin$ and
$\sigma_{\rm min} \equiv \sigma(\mmin)$.  We require that $\deriv
\fnew/\deriv z \le \deriv \fcoll/\deriv z$; the approximations in
equation (\ref{eq:dfnew}) violate this requirement at sufficiently
high redshifts.  The solid line associates $\mmin$ with a virial
temperature $T_{\rm vir}=10^4 \kel$.  The dotted and dashed lines
assume values of $\mmin$ an order of magnitude larger and smaller than
this value, respectively.  The figure reveals two important points.
First, $\deriv \fnew/\deriv z$ evolves slowly with redshift.  It does
not approach zero until $z \sim 6$, by which point reionization is
complete.  This is because $\mmin$ is either larger than or close to
the nonlinear mass scale at these early redshifts, so pristine
galaxies are still forming.  In fact, the magnitude of $\deriv
\fnew/\deriv z$ actually increases with cosmic time; its relative
importance decreases only because the number of halos \emph{larger}
than the threshold increases even more rapidly (as they correspond to
a higher value of $\nu$), and each of these objects continues to
accrete mass.  Second, the fraction of mass entering new objects
\emph{increases} with increasing $\mmin$ because such objects become
rarer and it takes the nonlinear mass scale longer to reach a larger
threshold.  On the other hand, $\deriv \fnew/\deriv z$ decreases
somewhat more rapidly in this case.

Because equation (\ref{eq:dfnew}) is only approximate, we describe an
analogous calculation with the extended Press-Schechter formalism
\citep{bond91,lacey} in the Appendix.  We show the results in the
bottom panel of Figure~\ref{fig:fdot}: in this calculation the
transition between the two modes is even slower (see also
\citealt{scann03}).  Thus a slow transition seems inevitable, and we
will clearly require winds to introduce sharp features in the
reionization history.

\subsection{Wind Model}
\label{windmodel}

Galactic winds are the most likely agent for spreading metals through
the IGM (e.g., \citealt{aguirre-full,madau01,scann02,furl-metals}) and
for accelerating the Pop III/Pop II transition. In this section we
will outline a model describing them.  In order to keep our methods as
straightforward as possible, we will examine a simple model for the
wind expansion that captures the expected scaling laws and calibrate
it against more advanced treatments.

The key input is the underlying energy budget of the winds.  The
available energy from supernovae is $W_{SN} = f_\star E_{SN}
m_b/\omega_{SN}$, where $f_\star$ is the star formation efficiency
(which may depend on halo mass), $E_{SN} = 10^{51} E_{51}$ ergs is the
energy per supernova, $\omega_{SN}$ is the mass in stars per supernova
event, and $m_b$ is the baryonic halo mass.  We fix $E_{SN}$ and
$\omega_{SN}$ here for simplicity.  For Pop II stars, we take $E_{51}
= 1$ and $\omega_{SN} = 100 \Msun$, in accord with a \citet{scalo} IMF
for stellar masses in the range $0.1$--$100 \Msun$.  For Pop III
stars, we take $E_{51} = 10$ and $\omega_{SN} = 100 \Msun$, assuming
that most of the stellar mass undergoes pair-instability supernovae
and maximizing enrichment from the first generation \citep{heger02}.

The simplest possible wind model assumes an energy-conserving point explosion in a constant density medium.  This well-known \citet{sedov} solution predicts that $R \propto (W_{SN} t^2/\rho)^{1/5}$, where $R$ is the radius, $\rho$ is the density, and $t$ is the elapsed time.  Although this solution neglects a number of important processes (see \S \ref{wind-caveat}), \citet{furl-metals} showed that the scaling remains approximately correct, although the normalization does change significantly.  Then we can express $\eta$, the ratio of total mass enriched by the wind to the mass of each galaxy, as
\bq 
\eta(m) = 27 \, K_w f_\star^{3/5} \ E_{51}^{3/5} \ \left( \frac{m}{10^{10} \Msun} \right)^{-2/5} \ \left( \frac{10}{1+z}
\right)^{3/5},
\label{eq:xim}
\eq 
where $K_w$ is a normalization constant that accounts for the many factors we have neglected (note that it could also include variations in $\omega_{SN}$).\footnote{In the following, we actually use $1 + \eta(m)$ to include self-enrichment of the galaxy.}  We have assumed that  $t \approx 1/3H(z)$ (i.e., each wind has propagated for half of the age of the universe).  Comparing to the results of \citet{furl-metals}, this solution yields radii 2--3 times too large, depending on the mass and redshift, suggesting that typical wind models have $K_w \sim {1\over 27}$--${1\over 8} f_w^{3/5}$, where $f_w$ accounts for radiative losses.  Of all possible cooling mechanisms, we include only Compton cooling by setting $f_w = {\rm min}(1,t_{\rm comp}/t_H)$, where $t_{\rm comp}$ is the Compton cooling time and $t_H \approx H^{-1}(z)$.  We ignore other possibilities (most importantly those in the dense interstellar medium of the host) in order to maximize enrichment.  

Note that the two stellar populations have different efficiencies
$\eta$.  In evaluating equation (\ref{eq:xim}) for a particular
galaxy, we assume that the fraction of Pop III stars is
$\fnew/\fcoll$, where $\fnew$ is the total mass fraction in Pop III
stars integrated over time.  This does slightly underestimate the
effects of Pop III stars because smaller halos have more Pop III stars
and are more efficient polluters.  However, a more detailed treatment
would require fully self-consistent merger rates, which are currently
unavailable.

\subsubsection{Caveats}
\label{wind-caveat}

Our simple model neglects a number of important effects, which we list
here for completeness.  We emphasize that \citet{furl-metals} included
all of these and showed that they do not affect the scaling.  The
first is radiative cooling, which we have already discussed.  Second,
the Sedov solution neglects the gravitational potential of the host
halo.  At large masses, gravity tends to flatten the mass dependence
\citep{furl-metals}; however, such halos are rare at these high
redshifts.  A third difficulty is in the choice of elapsed time,
because halos have a range of star formation histories.  Crudely, we
would expect more massive halos to be older, which steepens the mass
dependence.  On the other hand, the Sedov scaling also neglects the
evolving background density (and the halo density profile). Older
halos spend more time expanding into dense environments, which
flattens the mass dependence.  Finally, we ignored the fact that the
wind travels through an expanding medium, so that it need not
accelerate swept up material from rest.

\citet{furl-bigm} noted that we can set an upper limit to the wind
size by balancing the energy input with the energy required to
accelerate the enclosed baryons to the Hubble velocity at the outer
edge of the wind.  This procedure yields $\eta_{\rm max} \approx 2.2
\eta$ with exactly the same scalings as in equation (\ref{eq:xim}).
Thus the above estimate is reasonable; it is significantly smaller
than the maximal limit because the winds have not completed their
expansion.

\subsection{The Enrichment History}
\label{bias}

We now wish to compute the probability that a collapsing halo (with $m \approx \mmin$) forms in a region already enriched by galactic winds.  As a first step, given the enrichment efficiency $\eta(m)$, we can estimate the fraction of space with metals:
\bq
Q'_e(z) = \int_{\mmin}^{\infty} \deriv m \ \left( \frac{m}{\bar{\rho}}
\right) \ \eta(m) n(m). 
\label{eq:Qpe}
\eq 
This equation would be accurate if the galactic winds did not overlap or if their volumes were additive (as is the case with \hii regions, which occupy a net volume dictated by the total number of ionizing
photons).  Because winds expand at much less than the speed of light, the latter is not a good approximation.  If the host galaxies were randomly distributed, then the true filling factor would be $p'_e = 1 - \exp(-Q'_e)$.  

However, rather than the simple volume fraction of enriched material, we
actually wish to compute the probability that a new halo forms in an
enriched region .  Collapsing halos are biased and therefore lie near
existing halos (and their winds).  We will now estimate the importance of
this effect.  We first define $R_w$ to be the average wind size (in
comoving units).  The excess probability that two galaxies sit near each
other is parameterized by the correlation function $\xi_{gg}$, which to
linear order can be written as $\xi_{gg} = b_1 b_2 \xi_{\delta \delta}$, where
$b_1$ and $b_2$ are the halo biases \citep{mo96} and $\xi_{\delta \delta}$
is the dark matter correlation function.  For simplicity, we will use the
linear power spectrum for the latter, with the transfer function of
\citet{eis98}.

The probability that a newly-forming halo lies within $R_w$ of an existing galaxy is then approximately
\bq
Q_e =  Q'_e [1 + b_{\rm new} \bar{b}_w \xi_{\delta \delta}(R_w) ],
\label{eq:qe}
\eq 
where $\bar{b}_w$ is the mean bias of a wind, and $b_{\rm new}$ is the bias of a newly-formed halo.\footnote{Taking the average $\xi_{\delta \delta}$ over the wind volume does not significantly affect our results.} The mean bias of the enriched regions is:
\bq 
\bar{b}_w \equiv \frac{\int \deriv m \, m \, \eta(m)
\, b(m) \, n(m)} {\int \deriv m \, m \, n(m)},
\label{eq:bw}
\eq 
where the extra factor of $m$ enters because we compute the bias of the enriched \emph{volumes} rather than the bias of halos.  The bias of newly-formed galaxies is $b_{\rm new} \approx b(\mmin)$ because all such objects have mass close to the low-mass threshold for star formation.  Finally, we assume randomly distributed wind hosts by setting the enrichment probability $p_e=1-\exp(-Q_e)$.

\begin{figure}
\plotone{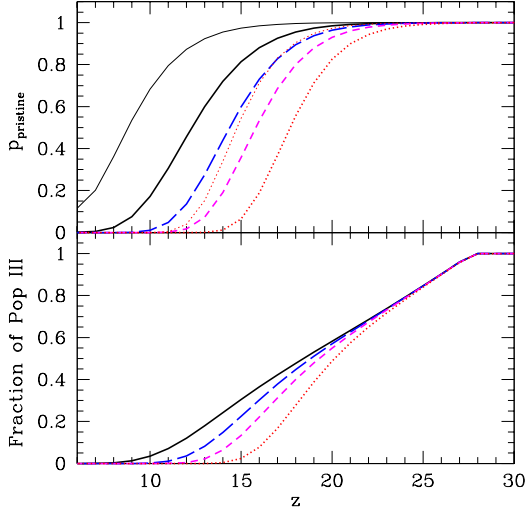}
\caption{\emph{Top:} Probability that a new halo forms out of pristine gas.  The thick curves include clustering, while the thin curves neglect it.  \emph{Bottom:}  The total fraction of collapsing gas forming Pop III stars; all curves include clustering. We take $f_\star^{\rm II}=f_\star^{\rm III} = 0.1$.   The solid, long-dashed, short-dashed, and dotted curves assume $K_w^{1/3} = 1/3,\,1/2,\,2/3$, and $1$, respectively.  }
\label{fig:penrdef}
\end{figure}

Figure~\ref{fig:penrdef} shows the resulting enrichment histories.  We
take $f_\star^{\rm II}=f_\star^{\rm III}=0.1$; the different curves
take different wind efficiency parameters $K_w$.  The thick curves in
the top panel show $p_{\rm pristine} \equiv 1 - p_e$, the probability
that a new halo forms Pop III stars.  The time at which Pop III stars
disappear obviously depends strongly on the wind efficiency parameter.
For realistic choices ($K_w \ll 1$), pristine halos become rare at $z
\la 10$.  The thin curves (shown only in the $K_w^{1/3}=1/3$ and 1
cases) illustrate the behavior if we neglect clustering: because
galaxies are so highly biased at these large redshifts, clustering
makes enrichment of galaxies much easier.  It completes when only
about half of the total volume has been injected with metals.  The
bottom panel shows the the total fraction of collapsing gas able to
form Pop III stars, $p_{\rm pristine} \times (\deriv \fnew/\deriv
z)/(\deriv \fcoll/\deriv z)$.  Although $Q_e'$ evolves fairly rapidly,
including the fraction of gas that accretes onto existing halos
significantly moderates the rate of decline.  In this model, the
evolution of $\deriv \fnew/\deriv z$ dominates the Pop III/Pop II
transition until relatively late, regardless of $K_w$.

\subsubsection{Caveats}
\label{enrich-caveat}

We now outline and justify the simplifications that led to equation
(\ref{eq:qe}).  The most serious shortcoming is our treatment of wind
overlap.  We have made no attempt to follow the dynamics of colliding
winds.  However, since the wind volume scales with the deposited
energy more slowly than linearly, combining their energies through
overlap will only reduce the total enriched volume relative to a case
without overlap.  Thus the net effect of overlap is to \emph{slow} the
transition from Pop III to Pop II and therefore strengthen our
conclusions in \S \ref{reion}.  Second, we have assumed that galaxy
formation in the wind-enriched regions can be well-described by dark
matter dynamics.  This may not be a good assumption if the winds
disturb a large fraction of the IGM by, for example, sweeping it into
a shell, stripping nearby halos as they collapse, or triggering star
formation behind the termination shock \citep{scann01,cen03-shock}.
If the feedback is positive, double reionization would become even
more difficult, but if the winds suppress galaxy formation the
prospects for double reionization would be enhanced.  (That case would
be qualitatively similar to photoheating, except that the feedback
would operate much more slowly.)

There is also a set of issues related to our simplified global
treatment of enrichment.  Equation (\ref{eq:qe}) only includes linear
halo bias, and we only use the linear $\xi_{\delta \delta}$.  The
latter is probably not a bad approximation, because small-scale
nonlinearities are primarily due to the density profiles of individual
halos (e.g., the halo model; \citealt{cooray02}).  In our case, all
the baryons inside each halo likely collect into a single galaxy, in
which case the one-halo nonlinearities are not relevant.
\citet{scann03} constructed a more detailed enrichment model that
included nonlinear bias.  They used the two-point halo mass function
\citep{scann-bark02} to compute the probability that a halo lies
within the winds of its neighbors.  Our extended enrichment histories
are qualitatively consistent with theirs, so the nonlinear corrections
included by this formalism do not seem to affect our conclusions.
Another crucial simplification is that we have taken the wind sources
to be randomly distributed.  In reality, the bias of the wind hosts
will increase the amount of overlap, decrease $Q_e$, and extend the
transition; in this sense our conclusions about double reionization
are conservative.  All of these issues are best addressed with
numerical simulations, which we forego in order to keep our models as
simple as possible.

\subsection{The Reionization History}
\label{reion}

We are now in a position to determine how enrichment, and the
accompanying transition from Pop III to Pop II star formation, affects
reionization.  The rates at which Pop III and Pop II stars form are
\bq
\frac{\deriv f_{\rm III}}{\deriv z} = p_{\rm pristine}(z) \,
\frac{\deriv \fnew}{\deriv z} 
\label{eq:dfpop3}
\eq
and
\bq
\frac{\deriv f_{\rm II}}{\deriv z} = \frac{\deriv \fcoll}{\deriv z} -
\frac{\deriv f_{\rm III}}{\deriv z}.
\label{eq:dfpop2}
\eq 
To translate these into production rates of ionizing photons, we
define $\zeta_{i}$ as the number of ionizing photons produced per
collapsed baryon for Pop $i$ (where $i=$II or III), e.g. $\zeta_i =
A_{\rm He} f_\star^i f_{\rm esc}^i N_{\gamma b}^i$, with $f_{\rm
esc}$ the fraction of ionizing photons able to escape the host
halo, $N_{\gamma b}$ the number of ionizing photons produced per
baryon incorporated into stars, and $A_{\rm He}$ a correction for helium.
We will fix $N_{\gamma b}^{\rm II}=4000$ (appropriate for a Salpeter
IMF) and $N_{\gamma b}^{\rm III}=30,000$ (appropriate for very massive
stars; \citealt{bromm-vms}).  Note that $f_\star$ affects both $\eta$
and $\zeta$, while $f_{\rm esc}$ and $K_w$ act as separate efficiency
parameters for the two mechanisms.  At least crudely, these are the
source parameters that we can tune in order to test the plausibility
of double reionization.

The number of ionizing photons emitted per baryon per unit redshift is then 
\bq 
\epsilon(z) =
\zeta_{\rm II} \frac{\deriv f_{\rm II}}{\deriv z} + \zeta_{\rm III} \frac{\deriv f_{\rm III}}{\deriv z},
\label{eq:eps}
\eq 
and the global ionized fraction $\bxi$ evolves according to \citep{shapiro87,barkana01} 
\bq
\frac{\deriv \bxi}{\deriv z} = \epsilon(z) - \bxi \alpha_B n_e C (1+z)^3 \left| \frac{\deriv t}{\deriv z} \right|,
\label{eq:xievol}
\eq 
where $\alpha_B$ is the case-B recombination coefficient, $n_e$ is the
comoving electron density in a fully ionized medium, and $C$ is the
density clumping factor within ionized regions.  The first term is the
rate at which stars produce ionizing photons, while the second is the
rate at which protons and electrons recombine.  We treat $C$ as a
constant; a better model would include the evolving density
distribution of ionized gas in the IGM.  \citet{miralda00} have
developed such a model calibrated to numerical simulations.  They
argue that low-density regions, where the recombination rate is small,
will be ionized first.  As the ionized volume increases, the ionizing
background can keep progressively denser regions ionized.  They showed
that while $C$ increases throughout reionization, it does so only
modestly until $\bxi \ga 0.9$, because nearly all of the volume is
near the mean density at the high redshifts of interest.  Since we are
primarily interested in the middle phases of reionization, the
assumption of a constant $C$ is reasonable for most of reionization
unless there is considerably more small-scale structure than the
simulations of \citet{miralda00} found (e.g., \citealt{haiman01-mh}).
However, it will underestimate the recombination rate at the end of
reionization.  (Note also that with a constant $C$ we must explicitly
force $\bxi \le 1$.)

\begin{figure}
\plotone{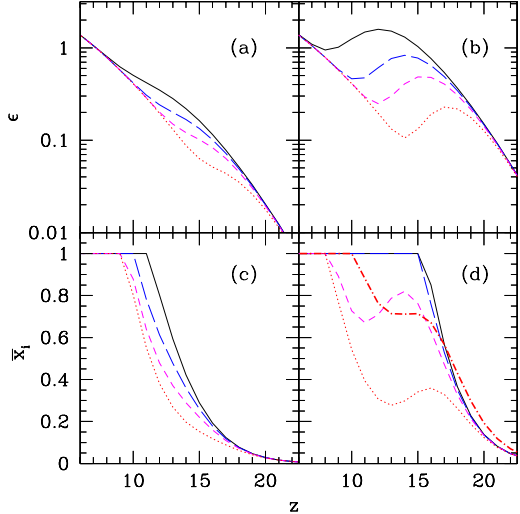}
\caption{Reionization histories.  \emph{(a):} Normalized emissivity in our fiducial model (with $f_\star^{\rm II}=f_\star^{\rm III}=0.1$ and $f_{\rm esc}^{\rm II}=f_{\rm esc}^{\rm III}=0.05$).  The solid, long-dashed, short-dashed, and dotted curves assume $K_w^{1/3} = 1/3,\,1/2,\,2/3$, and $1$, respectively.  \emph{(b):}  Same as \emph{(a)}, except $f_{\rm esc}^{\rm III}=0.4$.  \emph{(c):} Ionized fraction in the fiducial model, assuming $C=1$.  \emph{(d):}  Ionized fraction for the models in \emph{(b)}.  All curves assume $C=3$ except the dot-dashed line, which has $C=1$ but is otherwise the same as the dotted curve.  The corresponding enrichment histories are shown in Figure~\ref{fig:penrdef}. }
\label{fig:xbardef}
\end{figure}

Figure~\ref{fig:xbardef} plots the reionization history in two sets of
models.  The top panels show the evolution of $\epsilon(z)$ and the
bottom panels show $\bxi(z)$.  We vary the wind efficiency from
$K_w^{1/3}=1$ to $K_w^{1/3}=1/3$ in each panel.  In all cases, we
assume $f_\star^{\rm II}=f_\star^{\rm III}=0.1$ as in
Figure~\ref{fig:penrdef}; because these quantities affect both
enrichment and reionization they have little effect on our
conclusions.  Panels \emph{(a)} and \emph{(c)} show a baseline case in
which $f_{\rm esc}^{\rm II}=f_{\rm esc}^{\rm III}=0.05$.  The
difference in emissivity between the two generations is then
determined purely through the stellar physics encapsulated by
$N_{\gamma b}$.  In panels \emph{(b)} and \emph{(d)}, we set $f_{\rm
esc}^{\rm III}=0.4$ in order to exaggerate the difference between the
generations (although there is some motivation for a high escape
fraction from the first generation of sources; \citealt{whalen04}).
In panel \emph{(c)}, we assume that $C=1$, as is appropriate for the
bulk of the IGM according to \citet{miralda00}.  In panel \emph{(d)},
however, we assume $C=3$.

According to panel \emph{(a)}, the difference in $N_{\gamma b}$ between Pop
II and Pop III is clearly insufficient for double reionization: the slow
evolution in $\deriv \fnew/\deriv z$ smooths out the expected
discontinuity.  Not surprisingly, $\bxi$ is therefore monotonic regardless
of the wind efficiency.  Interestingly, this is true even though we have
``tuned'' the parameters so as to force complete enrichment and
reionization to occur at approximately the same epoch.  Thus, to introduce
features into $\bxi$ we must exaggerate the contrast between the
generations, as in panels \emph{(b)} and \emph{(d)}.  By manipulating the
escape fractions, we cause a clear drop in the emissivity during the Pop
III/Pop II transition.  However, even in this case we still only force
double reionization if the winds are powerful \emph{and} clumping is
strong.  We remind the reader that even with $f_w=1$, the wind enrichment
models of \citet{furl-metals} had $K_w \la 1/8$, so we regard larger values
as implausible.  Double reionization is difficult because
\emph{maintaining} full ionization requires only that the production rate
of ionizing photons exceeds the rate at which hydrogen atoms recombine
(i.e., the right-hand side of equation (\ref{eq:xievol}) must be positive).
Thus $\bxi$ will only decrease with time if \bq \epsilon(z) \la 0.1 \ C \
\bxi \ \left( \frac{1+z}{10} \right)^{1/2}.
\label{eq:doublecond}
\eq 
The marked decrease for the solid and long-dashed curves is large in
relative terms but still leaves the absolute emissivity strong enough
to cancel recombinations.  This illustrates an obvious feature of
these models: if reionization completes well before enrichment, the
emissivity will have risen sufficiently between these two epochs so
that, even after a sharp fall in $\epsilon$, the universe remains
completely ionized.  If we accelerate enrichment through more powerful
winds (the short-dashed and dotted curves), $\epsilon$ does fall below
the threshold of equation (\ref{eq:doublecond}), and a brief epoch of
recombination can ensue.  However, this is only true if $C$ is larger
than unity.  The dot-dashed curve shows $\bxi$ for the same model as
the dotted curve (i.e., maximal winds) but with $C=1$.  In this case,
$\bxi$ flattens but does not decline significantly.

We have explored a range of parameter choices and found that double
reionization can in principle be achieved, but only in rather extreme
models.  In order for the emissivity evolution to have a strong
feature, we require $\zthree \gg \ztwo$ (and in particular much larger
than the expected difference in $N_{\gamma b}$).  Second, enrichment
must be timed to occur during or soon after reionization, so that the
emissivity has not risen too far before enrichment is complete.
Third, the magnitude of the emissivity after the Pop III/Pop II
transition must satisfy equation (\ref{eq:doublecond}).  This is
easiest to achieve by boosting the IGM clumping factor, because that
accelerates recombinations and gives the emissivity less time to
recover.  We find that double reionization \emph{only} occurs in parts
of the IGM with $C \ga 3$.  Because such regions fill only a small
fraction of the volume in standard models (e.g., \citealt{miralda00}),
a recombination phase is strongly disfavored.

\subsubsection{Quasars and a Variable $f_\star$}
\label{varfstar}

There is some evidence that $f_\star \propto \sigma^2 \propto m^{2/3}
(1+z)$ in local galaxies with velocity dispersions $\sigma < \sigma_c
\equiv 124 \kms$ \citep{dekel03,kauffmann03}; above this threshold the star
formation efficiency remains constant.  How would such behavior in
high-redshift galaxies affect our conclusions?  We let $f_{\star c} \equiv
f_\star(\sigma_c)$ and assume $f_\star \propto m^{2/3}$ for all masses.
(Massive halos with $\sigma>\sigma_c$ are so rare at high redshifts that
they have a negligible effect on our results.)  Note that the Pop III star
formation efficiency is a constant because such stars form only in halos
with $m \sim \mmin$.  We can repeat our reionization calculations using
this prescription; the only adjustment is that we must modify equation
(\ref{eq:eps}) to account for the increasing $f_\star$ as galaxies grow
(e.g., \citealt{wyithe03}).  For a fixed energy input, the $f_\star \propto
m^{2/3}$ models tend to have somewhat slower enrichment because $\eta$ is
suppressed in the (abundant) low-mass halos in which winds are most
efficient (eq. [\ref{eq:xim}]).  It is also somewhat harder to achieve a
sharp drop in the emissivity because the typical galaxy mass increases with
cosmic time (and hence so does the mean $f_\star$).  However, the $f_\star$
law has no real effect on our conclusions regarding double reionization:
our direct calculations for the case of $f_\star \propto \sigma^2$ imply
that the same fairly extreme conditions outlined above must be fulfilled.
Thus, for the sake of brevity, we have chosen not to show any explicit
results for these models.

We note here one interesting result of our models: if $f_\star \propto
m^{2/3}$, early reionization at $z \gg 6$ requires the Pop III star
formation efficiency to be much larger than the Pop II efficiency in halos
of a comparable mass.  The reason is simply that Pop III stars must form in
pristine halos with $m=\mmin$, where $f_\star^{\rm II} \approx 0.01
f_{\star c}^{\rm II}$.  The number of Pop III stars would therefore also be
extremely small, unless $f_\star^{\rm III}(\mmin) \gg f_\star^{\rm
II}(\mmin)$.  Previous studies, such as \citet{wyithe03}, had found such
early reionization to be possible because they allowed more massive halos
(where the star formation efficiency is large) to form Pop III stars even
though such galaxies would have been self-enriched.  Thus, if observations
continue to favor early reionization, it appears the star formation
efficiency must scale differently with mass than in nearby galaxies, at
least for one of the two star formation modes.

Similar models could also describe the input of black holes, which may generate powerful winds (see \citealt{furl-bigm} and references therein).  \citet{wyithe03c} find a good fit to the quasar luminosity function over the range $z=2$--$6$ by assuming that $M_{\rm bh} \approx 10^5 \Msun (\sigma/54 \ {\rm km \ s}^{-1})^{5}$.  We can estimate how quasars affect double reionization by assuming that this relation extends to the higher redshifts and lower masses of interest.  We will also assume that a quasar radiates a fraction $\epsilon_Q$ of its rest mass and that a fraction $q_w$ powers the wind.  So long as $f_\star=$constant, quasars make only a small difference to the results if $\epsilon_Q \sim q_w \sim 0.1$ because black holes can be ignored in the small halos responsible for most of the enrichment and ionizing photons.  On the other hand, if $f_\star \propto m^{2/3}$, quasars can become significant in small halos.   For example, the ratio of wind input energies is
\bq 
\frac{W_{SN}}{W_{Q}} \approx 0.1 \
\left( \frac{f_{\star c}}{0.1} \ \frac{0.1}{q_w} \ \frac{0.1}{\epsilon_Q} \ E_{51}\right) \ \left( \frac{10}{1+z} \right)^{5/2}.
\label{eq:qsowind}
\eq
Quasars can be inserted directly into our $f_\star \propto m^{2/3}$ models without modification; clearly they will not alter the likelihood of double reionization, although they could alleviate the problem mentioned in the previous paragraph (but see \citealt{dijkstra04b}).

\section{Photoionization Heating}
\label{rad}

The second type of feedback is radiative, for which the most important
example is photoionization heating.  As described in \S \ref{intro},
this raises the Jeans mass in ionized regions by an uncertain amount
and suppresses accretion onto existing small halos.  Our primary
interest is in whether this feedback mechanism can cause double
reionization; we will therefore take a phenomenological approach and
leave the minimum galaxy mass in heated regions as a free parameter.
Because the physics depends on the depth of the potential well, we
will phrase the threshold in terms of a (redshift-independent) virial
temperature $T_h$.  We will also let $T_c = T_{\rm vir}(\mmin)$; the
\emph{h} and \emph{c} subscripts refer to hot and cold gas,
respectively.  We will take $T_h = 2.5 \times 10^5 \kel$ as a fiducial
value, corresponding to $V_c \approx 50 \kms$.  Once a region is
ionized, it will begin to cool adiabatically (and radiatively if it
recombines).  This will soften the increase in the time-averaged Jeans
mass (or ``filter mass''; \citealt{gnedin98}).  \citet{cen03} has
included these effects more precisely through a ``phase-space''
description.  By ignoring this possibility, we will maximize the
possibilities for double reionization.

A number of authors have already included photoionization heating in models of reionization (e.g., \citealt{wyithe03,cen03,somerville03,onken04}).  In this case the emissivity has two terms: 
\bq 
\epsilon(z) = x_h \frac{\deriv \langle \zeta f_h \rangle}{\deriv z} + (1 - x_h)
\frac{\deriv \langle \zeta f_c \rangle}{\deriv z},
\label{eq:epsmass}
\eq
where $x_h$ is the filling factor of regions that have been heated, $f_h \equiv \fcoll(>T_h)$, and $f_c \equiv \fcoll(>T_c)$.  We take the derivative of the mass-averaged ionizing rate for each component, $\langle \zeta f_h \rangle$ and $\langle \zeta f_c \rangle$, rather than of the simple collapse fractions to include the possibility that the star formation efficiency is a function of halo mass.  (The functional dependence of $f_\star$ qualitatively affects the results for photoionization heating, unlike in the metal enrichment scenario.) We may then insert this emissivity into equation (\ref{eq:xievol}) and solve for $\bxi(z)$.  Note that we assume both kinds of sources to have the same $\zeta$.

However, to do so we must specify how $x_h(z)$ evolves.  The simplest assumption, and one often used in the literature, is to set $\bxi=x_h$: the fraction of gas that has been heated is the same as the ionized fraction.  If this prescription is accurate, double reionization from radiative feedback is \emph{impossible even in principle}, because $\bxi$ regulates the feedback and there is no possibility of ``overshoot."  However, we must actually have $x_h \ge \bxi$ because some photons are lost to recombinations.  For example, suppose $T_h \gg T_c$, so that the emissivity in hot regions is negligible.  The first halos (with $T_{\rm vir} = T_c$) collapse and ionize their surroundings; structure formation effectively stops in these
ionized bubbles.  However, small halos can continue to form in the neutral gas outside these bubbles.  Thus ionizing photons will preferentially appear in \emph{cold} regions while recombinations will only occur in \emph{hot} regions.  The filling factor of hot gas must therefore evolve faster than $\bxi$.  The extreme limit would be to calculate $x_h$ without including any recombinations, 
\bq 
\frac{\deriv x_h}{\deriv z} = \epsilon(z).
\label{eq:xTevol}
\eq 
Of course, reality lies somewhere between these two extremes: some
fraction of the ionizing photons are produced in pre-ionized regions
and counteract recombinations before ionizing new material, while
recombinations also occur in the newly-formed ionized bubbles.  The
complexity of radiative transfer and feedback makes a self-consistent
choice difficult and is probably best addressed with simulations.  We
choose equation (\ref{eq:xTevol}) in order to maximize the
possibilities for double reionization; even in this extreme case we
will find double reionization to be quite difficult.

\begin{figure}
\plotone{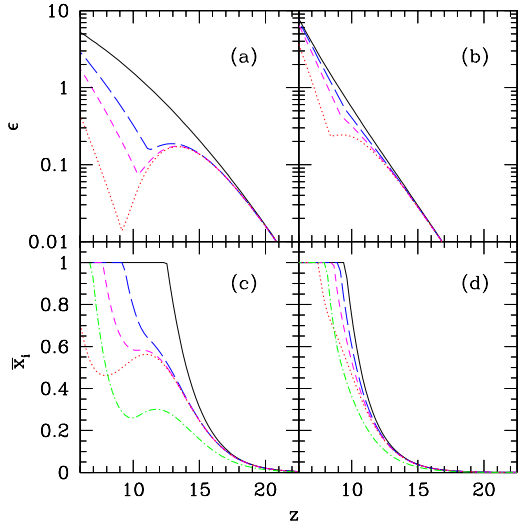}
\caption{Same as Fig.~\ref{fig:xbardef}, but for models in which photoionization heating raises the mass threshold for collapse.  Panels \emph{(a)} and \emph{(c)} assume $f_\star=$constant and $\zeta=100$, while panels \emph{(b)} and \emph{(d)} assume $f_\star \propto m^{2/3}$ and $\zeta_c=500$.  The solid, long-dashed, short-dashed, and dotted curves have $T_h=10^4,\, 10^5,\, 2.5 \times 10^5$, and $10^6 \kel$, respectively, with $C=1$. The dot-dashed curves have $T_h=2.5 \times 10^5 \kel$ and $C=3$; in this case the emissivities are the same as the short-dashed curves.}
\label{fig:xbarmS0}
\end{figure}

Figures~\ref{fig:xbarmS0}\emph{a} and \emph{c} show reionization histories for a set of models with $\zeta=100$ and $f_\star=$constant (note that we only consider a single stellar population here).  The solid, long-dashed, short-dashed, and dotted curves have $T_h=10^4,\, 10^5,\, 2.5 \times 10^5$, and $10^6 \kel$, respectively, with $C=1$.  The dot-dashed curve has $T_h=2.5 \times 10^5 \kel$ and $C=3$.  In this case, the emissivity is the same as for the short-dashed curve.  The solid curve shows the ionization history without photoheating; it rises steeply, completing reionization in a relatively short time period.  An increase in $T_h/T_c$ introduces a stronger break into the emissivity.  As long as $x_h \ll 1$, small halos dominate the ionizing photon budget and the evolution is nearly independent of $T_h/T_c$. Once $x_h$ becomes large, $\epsilon \propto \deriv \langle \zeta f_h \rangle/\deriv z$ and $T_h$ determines the amplitude as well as the duration of the break.  Note that the redshift at which $x_h=1$
corresponds to the minimum in $\epsilon$.  Unlike in the metal enrichment case, this \emph{must} occur while $\bxi<1$.  Thus, any feature in the ionization history due to photoionization will occur before reionization is complete.

For this reason, a sharp drop in the emissivity is more difficult to achieve,  because $\bxi$ (or in our model the closely related quantity $x_h$) regulates the emissivity and causes it to flatten substantially before it can drop.  Equation (\ref{eq:doublecond}) again expresses the criterion for double reionization: the emissivity must fall below the recombination rate.  Not surprisingly, it therefore requires either extremely large suppression ($T_h \sim 10^6 \kel$) or substantial clumping.  Our fiducial choice $T_h=2.5 \times 10^5 \kel$ is just on the threshold for achieving double reionization if $C=1$.

Figures~\ref{fig:xbarmS0}\emph{b} and \emph{d} show reionization histories for models with $f_\star \propto m^{2/3}$ and $\zeta_c=500$ (evaluated at the fiducial $\sigma_c$).  In this case the behavior is obviously quite different, with $\bxi$ remaining monotonic in all cases.  Even for the most extreme choice of $T_h/T_c$, $\epsilon(z)$ exhibits only a slight turnover.  Extended reionization histories are more difficult to achieve because the star formation efficiency in low-mass halos is small, making their contribution to the global emissivity low and their suppression much less significant.  The decline in emissivity is much less severe in this case, and double reionization requires truly extreme parameters.  This also implies that if quasars are important in reionization and $M_{\rm bh} \propto \sigma^5$, they will tend to wash out double reionization because massive halos contribute a large fraction of the ionizations.

We have not included halo bias in the feedback treatment in this section.  This is unlikely to affect our final results for two reasons.  First, the bias should be less important than for metal enrichment, because overlapping \ion{H}{2} regions grow quite large even during the middle stages of reionization (FZH04), spanning several comoving Mpc when $\bxi \sim 0.5$.  On these scales $\xi_{\delta \delta}$ is small and the ionized regions are probably close to a fair sample of the universe.  Because of this overlap, it is also extremely difficult to calculate the relevant scale from first principles (see \S \ref{bubbles} for a first look at the issues involved).  The second reason is that photoheating guarantees that bias will modulate ionizations and feedback in the same way.  In other words, if halos are overabundant in ionized regions, the emissivity will begin to decline at a smaller $\bxi$.  But both the heating and ionizing rates will decline together, so the net effect will simply be for feedback to flatten the emissivity sooner.  The bias will not be completely degenerate if we use equation (\ref{eq:xTevol}) but is still not likely to have a strong effect because that scenario essentially assumes that all ionizing photons come from widely separated (and hence unbiased) sources.  In any case, a full treatment of the interplay between bias and feedback will have to await self-consistent numerical simulations.

In summary, photoheating can in principle cause double reionization, but
only under the following restricted conditions: \emph{(i)} the suppression
of structure formation is more severe than current estimates ($T_h \gg
T_c$) and/or the universe is clumpy, \emph{(ii)} star formation is
efficient in small halos, and \emph{(iii)} nearly all ionizing photons go
into ionizing new material, rather than counteracting recombinations in
ionized regions (so that $x_h > \bxi$).  Taken together, we regard these
conditions as implausible, and conclude that double reionization is
unlikely to occur through photoheating.

\section{Photodissociation of H$_2$}
\label{h2}

For completeness, we now examine another radiative feedback mechanism.
As described briefly in \S \ref{intro}, gas in the first collapsed
halos cools through rotational transitions of H$_2$
\citep{haiman96,tegmark97}, which operate for $T_{\rm vir} \ga 200
\kel$.  However, such molecules are fragile and a weak soft UV background
of $J_{21} \ga 10^{-3}$ suffices to dissociate them, where $J_\nu
\equiv J_{21} \times \junits$ \citep{haiman00,machacek01}.  Crucially,
any photon with energy above 11.26 eV can dissociate H$_2$, so (unlike
in the case of ionizing photons) any individual halo sees Lyman-Werner
sources out to cosmological distances \citep{haiman97}.  The UV
background redward of the Lyman limit builds up quickly as stars form,
suppressing cooling via H$_2$ and increasing the minimum halo mass for
star formation from $T_c \sim 200 \kel$ to the atomic cooling
threshold of $T_h = 10^4 \kel$.

Because the effects of H$_2$ photodissociation can be phrased in a
similar way to photoheating (except that the ``heated'' volume is
automatically larger than the ionized volume), we can again find
scenarios in which $\bxi$ decreases with cosmic time, as long as the
clumping factor is substantial and $T_h/T_c \gg 1$.  However, it is
easy to show that this phase occurs early in the reionization process,
because the photodissociation threshold \emph{must} be reached well
before reionization is completed \citep{haiman00}.  We wish to
estimate $J_\nu$ in the Lyman-Werner bands; neglecting the line
opacity of the IGM, we have
\bq
J_\nu = \frac{c}{4\pi} \int_z^{z_{\rm max}} \deriv z' \ \left| \frac{\deriv
t}{\deriv z'} \right| \ \epsilon_{\rm LW}(\nu') \ \frac{1+z}{1+z'},
\label{eq:j21}
\eq 
where $\epsilon_\nu$ is the emissivity per unit frequency,
$\nu'=\nu(1+z')/(1+z)$, and the last factor accounts for the
cosmological redshifting of the photon energy.  The upper limit
$z_{\rm max}$ enters because ionizing photons cannot propagate through
the neutral IGM.  The Lyman series introduces a ``sawtooth
modulation'' of the background \citep{haiman97}, but this does not
affect our estimate.  To connect to the ionized fraction, we write
$\epsilon_{\rm LW}(\nu) = h \nu \times (1/\nu) \times \chi
\dot{n}_{\rm ion}$, where $\dot{n}_{\rm ion}$ is the rate at which
ionizing photons are produced per unit volume and $\chi$ is the number
of Lyman-Werner photons per frequency decade divided by the number of
ionizing photons per frequency decade.\footnote{We will assume for
concreteness that all of the ionizing photons are contained in the
frequency decade closest to the Lyman limit, which is a good
approximation for stellar spectra.  Even if the ionizing sources have
hard spectra similar to quasars, our conclusion would
not change.}  \citet{ciardi03} have evaluated a closely related
quantity and found $\chi \sim 10^{0.4}$--$10^{1.2}$ for typical
stellar populations (the only difference is that they calculated the
number of continuum photons just shortward of Ly$\alpha$ rather than
shortward of 11.26 eV).  Thus using our earlier definitions, we have
\bq 
J_\nu \approx \frac{c\,h}{4\pi}\ \chi\
\bar{n}_b(z)\ f_\star \ N_{\gamma b} \ \Delta \fcoll,
\label{eq:japprox}
\eq
where $\bar{n}_b(z)$ is the mean baryon density and $\Delta
\fcoll=\fcoll(z) - \fcoll(z_{\rm max})$.  Neglecting recombinations,
we can then write $\bxi 
= A_{\rm He} f_\star f_{\rm esc} N_{\gamma b} \fcoll$, so the ionized fraction
$\bar{x}_m$ when we reach the photodissociation threshold $J_{m,21}$
is
\bq
\bar{x}_m \approx 0.03 \  \left( \frac{f_{\rm esc}}{\chi} \right) \
\left( \frac{J_{m,21}}{0.1} \right) \ \left( \frac{\fcoll}{\Delta
  \fcoll} \right) \ \left( \frac{1+z}{10} \right)^3.  
\label{eq:xh2}
\eq
Even with $f_{\rm esc}=1$, $\chi=1$, and $\fcoll \approx 3 \Delta
\fcoll$, the ionized fraction is at most several percent when we reach
the photodissociation threshold.  This estimate agrees with the
detailed models of \citet{haiman00} (their Figure 7).  Thus although
this mechanism can cause $\bxi$ to turn over, it can only do so near
the beginning of reionization and the amplitude of the change
will be small.

Finally, we note that in reality the chemistry of H$_2$ is
sufficiently complicated that the detailed consequences of radiative
feedback remain unclear.  Initially it was thought that X-rays could
catalyze its formation by increasing the free electron fraction
\citep{haiman00}; however, \citet{oh03-entropy} showed that the
heating that inevitably accompanies X-rays impedes collapse, ensuring
a net suppression.  Others have argued for more complicated positive
feedback mechanisms near \hii regions \citep{ricotti,cen03}.  We will
ignore all of these possibilities simply because positive feedback
obviously cannot cause double reionization.  Moreover, the
``self-regulation'' process will not necessarily completely halt the
formation of low-mass halos.  If small halos are responsible for the
background, they will reach an equilibrium in which the UV background
prevents the smallest halos from forming stars but allows more massive
and better shielded halos to cool.  Again, such a scenario tends to
wash out double reionization.

\section{The Topology of \ion{H}{2} Regions}
\label{bubbles}

One promising observable of the reionization process is the size
distribution of \hii regions and its evolution with time (FZH04;
Wyithe \& Loeb 2004c), which can be probed through 21 cm tomography
\citep{furl04b} or through Ly$\alpha$ absorption
\citep{furl04c,wyithe04-bub}.  By modulating the ionizing efficiency
through local feedback mechanisms, the schemes we have described can
in principle have substantial effects on the size distribution of
ionized bubbles.  Here we will briefly discuss those consequences.
\citet{furl04b} made some progress in this direction by considering
simple scenarios with multiple generations of sources.  We will
improve on that treatment by examining the transition between star
formation modes in more detail.

We take the model of FZH04 as a starting point.  We assume as above
that the number of ionizing photons produced in a region is
proportional to the collapse fraction within the region.  We denote
the proportionality constant by $\zeta$.  Note that this differs from
the definition in \S \ref{reion} in two ways: it is cumulative over
the integrated star formation history and it includes a correction for
past recombinations.  A region can ionize itself if $\zeta \fcoll>1$.
FZH04 showed how to transform this simple condition into a size
distribution by including the implicit scale dependence of $\fcoll$
and ionizations from neighboring regions.  We rewrite the ionization
constraint as a condition on the density,
\bq 
\delta_m \ge \delta_x(m,z) \equiv \delta_c(z) - \sqrt{2} K(\zeta)
[\sigma_{\rm min}^2 - \sigma^2(m)]^{1/2},
\label{eq:barrier}
\eq 
where $K(\zeta) = {\rm erf}^{-1}(1 - \zeta^{-1})$.  Equation
(\ref{eq:barrier}) is then used as an absorbing barrier within the
excursion set formalism \citep{bond91,lacey} to construct the mass
function of ionized bubbles.  The crucial point is that the
\emph{shape} of the barrier determines the distribution of bubble
sizes.  The above barrier is nearly linear in $\sigma^2$, which
implies \citep{sheth98} a well-defined characteristic size for the
\hii regions.  Thus, the shape of the barrier is fixed by the
condition $\fcoll=$constant, while the normalization is fixed by the
ionizing efficiency.

We may now incorporate feedback by considering two sets of sources
with different ionizing efficiencies and/or different galaxy mass
thresholds.  In the latter case (i.e., photoionization heating), the
qualitative effects are easy to guess.  A high-density region forms
halos with $T_{\rm vir} \approx T_c$ which ionize the neighborhood.
Without feedback, the associated bubble would grow through two
processes: ({\it i}) it would merge with other \hii regions, and ({\it
ii}) halos inside it would continue to accrete rapidly, producing more
ionizing photons.  However, if $T_h \gg T_c$, structure formation is
suppressed in this region, the ``internal'' ionizations diminish, and
the bubble can grow only by merging with its neighbors.  Thus we would
expect \hii regions to be \emph{smaller} and more numerous, because
those around high-density regions grow more slowly and force voids to
be ionized by the relatively rare galaxies embedded inside.  At the
same time, the high-density dormant regions begin to recombine.
\citet{furl04b} argued that this process would not stop until the
entire universe had been ionized, though not necessarily
simultaneously (see \S \ref{rad}), so eventually the pattern induced
by recombinations (and exaggerated by the time delay between
reionization in different regions) would dominate the signal.
Unfortunately, because the photoheating depends directly on the
ionized fraction, it cannot be incorporated into the FZH04 model.

The FZH04 model can accommodate a transition due to metal
enrichment more easily, because the ionizing efficiency does not
depend on the local ionized fraction in this case.  We will assume
that there exists a function $\ptwo(\delta)$ characterizing the
probability that any source in a region of overdensity $\delta$
contains Pop II stars.  Physically plausible models will have
$\ptwo'>0$, where a prime denotes the derivative with respect to
density, because structure formation is more advanced in dense
regions.  Given the ionizing efficiency of each star formation mode,
the condition for self-ionization is
\bq 
\fcoll(\delta) \ [ (\ztwo-\zthree) \ptwo(\delta) + \zthree ] > 1.
\label{eq:pdcond}
\eq
This modified barrier has two effects on the bubble size distribution.
Unlike for a single type of ionizing sources, the left-hand side need
not be a monotonically increasing function of the overdensity
$\delta$.  Its derivative vanishes when
\bq
\ptwo + \frac{\fcoll}{f'_{\rm coll}} \ptwo ' =
\frac{\zthree}{\zthree-\ztwo} \ .
\label{eq:monoton}
\eq
Physically, if $\zthree \gg \ztwo$ and if high-density regions have
enriched themselves, then it is possible for these regions to remain
neutral even though regions with slightly smaller densities (and thus
incomplete enrichment) can self-ionize.  Obviously the increased recombination rate in dense regions will amplify this trend.

As an example, we consider the choice $\ptwo=1 - \exp(-\eta f_{\rm coll})$,
which essentially reflects the wind model from \S \ref{wind} (ignoring
$\fnew$).  Here the enrichment probability is a function of density through
its dependence on the collapse fraction; the exponential accounts for
overlap of bubbles from randomly distributed sources within each region.
Equation (\ref{eq:monoton}) then implies that the ionizing efficiency turns
over at $\eta \fcoll \approx 0.8$ for $\zthree \gg \ztwo$, and some regions
sufficiently far above this threshold could remain neutral.\footnote{Note
that $\fcoll$ actually depends on the scale of interest (as could $\ptwo$).
We neglect this here, considering only those scales for which $\sigma_{\rm
min} \gg \sigma(m)$.}  Note that not every region with $\eta \fcoll > 0.8$
would be neutral; above this point, the effective $\zeta$ decreases with
increasing $\delta$ but not necessarily by enough to become neutral.
Moreover, a dense region could still be ionized by its neighbors
\citep{barkana04}.  The turnover with density can mimic ``outside-in''
reionization in which voids are ionized before intergalactic sheets and
filaments (where the sources are located).  However, for this
effect to be significant we must have $\eta \fcoll(\delta=0) \approx 0.8$;
this only occurs when enrichment nears completion.  Thus in this simple, wind-driven
model we do not expect neutral dense regions to be common during
reionization.

Unfortunately, once the ionizing efficiency turns over, the excursion
set formalism of FZH04 breaks down.  This is because the number of
ionizing photons is no longer additive in a simple way, but rather
depends on the degree of enrichment of substructure inside the region
of interest.  However, if $\eta \fcoll(\delta=0)$ remains reasonably
small during reionization then the model can still be used.  In this
limit, we consider how the shape of the barrier -- and hence the shape
of the bubble size distribution -- changes due to the metal enrichment
pattern.  In standard reionization scenarios, the shape is set by
$\fcoll=$constant (the solution to $\zeta \fcoll=1$).  With
enrichment, equation (\ref{eq:pdcond}) shows that the shape is instead
set by a condition on a combination of $\fcoll$ and $\ptwo$.
Unfortunately, the dependence is weak for reasonable choices of
$\ptwo$.  For example, with the simple wind model examined above,
$\ptwo$ is purely a function of $\fcoll$ and the shape is still
determined by $\fcoll=$constant; it can be exactly reproduced by
choosing an effective $\zeta_{\rm eff}$ to match the total ionized
fraction.

\begin{figure}
\plotone{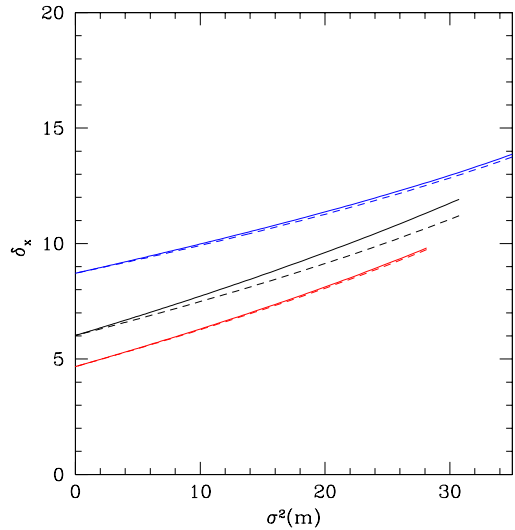}
\caption{Sample barriers from equation~(\ref{eq:ptwo}) (solid curves)
and for a single type of source (dashed curves) with the same limiting
value at $\sigma^2=0$.  In the solid curves, $\zthree=44$, $\ztwo=9$, and
$A=1$.  We show results for $\eta=5,\,15,$ and $25$, from bottom to
top. }
\label{fig:barriers}
\end{figure}

Thus, to change the shape, $\ptwo$ must have a more complicated
density dependence.  Let us consider the simple parameterization
\bq
\ptwo=1 - \exp \left[-\eta \fcoll \left( 1+A \frac{|\delta|}{1+z}
  \right) \right],
\label{eq:ptwo}
\eq 
where $A$ is some constant that measures the steepness of the density
dependence.  This form has the limits $\ptwo \rightarrow 1$ for
$\delta \rightarrow \infty$ and $\ptwo \rightarrow 0$ for $\delta
\rightarrow -\infty$, and it concentrates enrichment in dense regions
more than the simple wind model does.  One reason this may happen is
that $\fnew/\fcoll$ is smaller than the average in dense regions
because they are further along in the structure formation process.
Qualitatively, we expect the barrier to be somewhat steeper with this
prescription, because dense regions have extra enrichment and fewer
ionizing photons.  The solid curves in Figure~\ref{fig:barriers} show
some example barriers.  In each case, $\zthree=44$, $\ztwo=9$, and
$A=1$.  We show $\eta=5,\,15,$ and $25$, from bottom to top.  The
dashed curves show the barrier for a single source type with the value
at $\sigma^2=0$ fixed to match the solid curves.  These have
$\zeta_{\rm eff}=38,\,24$, and $11$ for the three values of $\eta$.
Indeed the barriers are steeper, which implies fewer small bubbles
because small, dense clusters are highly enriched.  However, we find
that the steepening is relatively weak, especially when one of the two
star formation modes dominates (i.e, $\zeta_{\rm eff} \approx \zthree$
or $\ztwo$).  Of course, if enrichment is sufficiently extensive that
equation (\ref{eq:monoton}) is violated, the barrier approach breaks
down and the bubble pattern could change significantly.

We have considered other possible parameterizations of $\ptwo$ based
around the simple wind expression; in the cases we examined, the
differences were not much larger than shown in
Figure~\ref{fig:barriers}.  The reason is that $\ptwo$ is simply used
to interpolate between two different populations, each of which would
follow the usual barrier.  In addition, the interpolation between them
also depends on the collapse fraction.  Thus it appears difficult to
change the shape dramatically, although exotic scenarios in which the
collapse fraction played no role could have stronger effects.

\section{Discussion}
\label{disc}

We have examined a variety of feedback mechanisms that could potentially
cause ``double reionization,'' in which the globally-averaged ionized
fraction $\bar{x}_i$ decreases over a limited period of cosmic time. These
mechanisms include metal enrichment, photoionization heating, and
photodissociation of H$_2$.  We constructed simple models that nevertheless
retained the crucial physics; as a result we were able to vary a relatively
small set of parameters to pin down the requirements for double
reionization.  In all cases, we found that double reionization requires
extreme (though not impossible) parameter choices.  In particular, it
requires a rapid drop in the ionizing emissivity over a single
recombination time (see equation [\ref{eq:doublecond}]).  Because the
recombination time varies spatially within the IGM, we found that it is not
difficult to imagine relatively dense pockets recombining after feedback
becomes strong (which could affect the topology of ionized gas; see \S
\ref{bubbles}), but because these occupy only a small fraction of the
universe, double reionization is difficult to arrange on large scales.

For metal enrichment, we require three conditions: ({\it i})
$\zthree/\ztwo \gg 1$, ({\it ii}) an average clumping factor
significantly larger than unity, and ({\it iii}) winds that expand
much more rapidly than predicted by existing semi-analytic models
\citep{madau01,scann02,furl-metals}.  We have argued that condition
({\it ii}) is implausible for the majority of the IGM
\citep{miralda00} and that ({\it iii}) is similarly implausible, so we
regard this mechanism as unlikely to cause double reionization.
Existing work predicted double reionization only because of an
artificial, instantaneous transition between star-forming modes
(\citealt{wyithe03,cen03}; see also \citealt{haiman03}).  Our models
are more similar to \citet{scann03}, who found a gradual disappearance
of Pop III stars.  In a hierarchical picture, the decline has two
causes: accretion shifts to higher mass, pre-enriched halos even when
the global metallicity is small and the slow expansion of galactic
winds.  The first depends on unknown halo merger rates, but our
estimate appears to be conservative.

Double reionization from photoheating is similarly difficult. In this
case we require ({\it i}) $T_h/T_c \gg 1$, ({\it ii}) small halos
(those with $T_{\rm vir} < T_h$) contribute a large fraction of the
ionizing photons, and ({\it iii}) new ionizing sources form in regions
that have not yet been ionized, so that their photons are not
``wasted" on recombinations (because in that case they do not heat the
IGM).  The second condition implies that the star formation efficiency
does not drop too rapidly with decreasing halo mass.  In particular,
if $f_\star \propto m^{2/3}$ persists at high redshifts (as has been
observed in nearby galaxies; \citealt{dekel03,kauffmann03}), then
small halos do not dominate the ionizing photon budget and double
reionization cannot occur.  Quasars also help to wash out double
reionization if the black hole mass scales steeply with host halo
mass, as in \citet{wyithe03c}.  We argued that the photoheating occurs
more rapidly than ionization, because photons are preferentially
emitted in regions that have remained neutral to that point, while
recombinations are confined to ionized regions.  We made an extreme
assumption in order to maximize the likelihood of double reionization;
this picture can be tested with self-consistent photoheating feedback
in simulations.  If $x_h \approx \bxi$ is more accurate, double
reionization is \emph{impossible} to achieve through photoheating.
Our fiducial choice for the degree of suppression, $T_h/T_c = 25$ is
on the threshold of double reionization; if photoheating is less
efficient at suppressing galaxy formation near the time of
reionization \citep{dijkstra04a}, then this mechanism also cannot
cause double reionization.  Note also that, if the first sources
produce X-ray photons (e.g., Ricotti et al. 2004; but see Dijkstra et
al. 2004b), the IGM will be heated gradually and (nearly) uniformly,
again suppressing accretion onto small halos \citep{oh03-entropy}.
This would tend to smooth out the transition from $T_c$ to $T_h$,
decreasing the likelihood of double reionization.  Clearly, double
reionization through either metal enrichment or photoheating requires
the confluence of a number of unlikely possibilities.

Finally, H$_2$ photodissociation has a similar phenomenological effect
to photoheating, increasing the minimum mass for star formation
(albeit from a much lower base level).  However, we showed that the
feedback threshold must be reached when the ionized fraction is still
small, so the recombination phase (even if it occurred) would be
difficult to observe.  \citet{haiman00} also reached this conclusion
through more detailed modeling.

We have not examined any combinations of these different mechanisms.
Together they could increase the contrast between the initial and
final ionizing efficiency, which would make double reionization
somewhat easier to achieve.  For example, photoionization heating
could extend reionization over a long interval.  This would increase
the time over which winds can expand and hence move the Pop III/II
transition closer to reionization, increasing the effective amount of
suppression in ionized regions.  Another possibility is that star
formation through H$_2$ cooling at extremely high redshifts could
begin to spread metals throughout the IGM.  Once photodissociation
dominates, there would be a long pause until reionization could
continue.  Again, the winds would have extra time to expand.  However,
we see no reason for any such scenario to increase qualitatively the
likelihood of double reionization.

We have focused here on particular ionization histories in which
$\bxi$ turns over.  We found that such scenarios are difficult to
achieve.  But we stress that \emph{extended} reionization is not
nearly so difficult; any form of feedback will help to prolong the
reionization era and to relieve the apparent tension between the $z
\sim 6$ quasar data \citep{fan,wyithe04-prox,mesinger04} and the
optical depth to electron scattering measured by \emph{WMAP}
\citep{kogut03}.  Indeed, a number of our models in
Figures~\ref{fig:xbardef} and \ref{fig:xbarmS0} show that feedback can
significantly extend the era over which $\bxi \sim 0.5$.  Moreover, we
have found that a turnover in the emissivity is also not difficult to
achieve; many of our feedback scenarios yield this kind of behavior,
at least if low-mass halos have high star formation efficiencies.
Unfortunately, the emissivity evolution is harder to measure.

Finally, we must also stress that we have studied double reionization
in a \emph{global} sense.  Although we have found that $\bxi$ only
turns over in exceptional circumstances, the reionization history of
any particular region or line of sight need not be monotonic.  For
example, if photoheating suppression is important, a volume ionized
early on will experience a sharp decrease in its emissivity and begin
to recombine.  Only later will halos grow sufficiently massive to
ionize it again.  Our models show that, while this region is
recombining, others are also being ionized, and the latter generally 
dominate the global evolution.  The most promising way to probe
reionization in different physical volumes is 21 cm tomography
\citep{mmr,zald04,furl04b}.  We have argued that these feedback
mechanisms should induce signatures in the bubble size distribution,
although metal enrichment appears to have little effect except in
dense, highly-enriched regions.  Because of the complexity of this
problem, numerical simulations would provide the best method to
examine changes in the bubble pattern in greater detail.

\appendix
\section{An Extended Press-Schechter Estimate for the Halo Formation Rate}
\label{eps}

In \S \ref{fnew}, we estimated $\deriv \fnew/\deriv z$ in a simple
fashion.  Here we point out that self-consistent analytic calculations
for this quantity do not yet exist.  Given the comoving number density
of halos per unit mass as a function of redshift, $n(m,z)$, and the
rate at which these halos merge per unit volume, $n(m_1,z) \ n(m_2,z)$
$Q(m_1,m_2,z)$, we can compute the rate at which halos merge to form
new objects with masses above $\mmin$.  Unfortunately, as recently
emphasized by \citet{benson04}, the merger rates typically used in
cosmological studies \citep{bond91,lacey} are not well-defined.  The
problem is most easily seen by noting that $Q(m_1,m_2) \neq
Q(m_2,m_1)$ in the extended Press-Schechter (EPS) formalism.  The
fundamental difficulty is that this formalism does not assign points
in space to unique halos: the smoothing procedure implicit in these
approaches requires that neighboring points be identified with halos
of different masses, even though they must physically be part of the
same object.  Merger rates, which by definition require uniquely
specified halos, are thus ill-defined quantities.  As a consequence,
merger tree algorithms based on the EPS formalism either do not
reproduce the proper mass function or do not conserve mass during the
individual time steps, although ad hoc procedures to alleviate these
problems have been developed (e.g., \citealt{somerville99}).

Nevertheless, it is useful to compute $\deriv \fnew/\deriv z$ with the
EPS formalism in order to investigate whether the slow evolution in
equation (\ref{eq:dfnew}) is an artifact of our approximation.  A
comparison to the EPS estimate is motivated by its succes in areas
such as semi-analytic galaxy formation (e.g.,
\citealt{kauffmann93,cole94,somerville99}) and modeling the quasar
luminosity function (e.g., \citealt{wyithe-lumfcn}).

For a halo of mass $m$ at redshift $z_l$, the EPS formalism provides
the fraction of matter that was in objects with masses smaller than
$\mmin$ at some earlier time $z_h$ \citep{lacey} and hence the
fraction of a halo's mass that was accreted recently:
\bq 
F(<\mmin,z_h|m,z_l) = {\rm erf}
\left[ \frac{\delta_c(z_h) - \delta_c(z_l)}{\sqrt{2(\sigma_{\rm min}^2 -
\sigma_m^2)}} \right].
\label{eq:discrete}
\eq 
The relevant timescale is the dynamical time within a galaxy (which
gives approximately the time over which star formation occurs),
$t_{\rm dyn} \sim (G \rho)^{-1/2} \sim (\Delta_v \rho_c G)^{-1/2} \sim
\sqrt{4/27\pi} H^{-1}(z)$, where $\rho_c=3H^2/8\pi G$ is the critical
density, $H(z)$ is the Hubble parameter at redshift $z$, and
$\Delta_v$ is the virial overdensity \citep{barkana01}.  If we then assume that
accretion onto halos with $m < 2 \mmin$ can be included in the
newly-formed halo component, we can estimate the fraction of
collapsing gas that accretes directly onto existing halos via
\bq 
f_{\rm old} = \frac{ \int_{2 \mmin}^\infty \deriv m \ m \ n(m) \
F(<\mmin,z_h|m,z_l) } { \int_{\mmin}^\infty \deriv m \ m \ n(m) \
F(<\mmin,z_h|m,z_l) },
\label{eq:fold}
\eq 
so that $\deriv \fnew/\deriv z \approx (1-f_{\rm old}) \deriv f_{\rm
coll}/ \deriv z$.  The denominator is necessary to normalize the total
mass accretion rate to its actual value; a direct calculation with the
EPS merger rates of the total collapse rate does \emph{not} reproduce
the true $\deriv \fcoll/\deriv z$.  One advantage of this formulation
is that it includes the duration over which a new halo must remain
isolated (unlike in equation [\ref{eq:dfnew}]).  In other words, a
halo that merges into a massive object immediately after passing above
$\mmin$ will most likely form Pop II, rather than Pop III, stars.
We have used the dynamical time to fix $z_h$.
Varying the time offset has only a small effect on the results:
increasing or decreasing $t_{\rm dyn}$ by a factor of two changes
$\deriv \fnew/\deriv z$ by $\la 10\%$ and leaves the shape unaffected.
Another approach is to use the explicit expression for $Q(m_1,m_2,z)$
in the EPS formalism to compute $\deriv \fnew/\deriv z$ directly (but
without the time offset).  This procedure (with symmetrized EPS merger
rates) yields results similar to equation (\ref{eq:fold}).

The resulting formation rate of new halos is shown in the bottom panel
of Figure~\ref{fig:fdot} for the same three mass thresholds as in the
top panel.  Interestingly, in this case $\deriv \fnew/\deriv z$
evolves even more slowly than our standard calculation predicts: it
falls below unity earlier and remains substantial much later.  We thus
conclude that $\deriv \fnew/\deriv z$, while unknown in detail, has
the important property that it evolves slowly and smoothly with
redshift, vanishing only at $z \la 6$ for the mass thresholds of
interest.  

\acknowledgements SRF thanks Z. Haiman, M. Kamionkowski, and S.~P. Oh
for helpful discussions.  This work was supported in part by NSF
grants AST-0204514, AST-0071019 and NASA grant NAG 5-13292 (for A.L.).


\begin{thebibliography}{75}
\expandafter\ifx\csname natexlab\endcsname\relax\def\natexlab#1{#1}\fi

\bibitem[{{Abel} {et~al.}(2002){Abel}, {Bryan}, \& {Norman}}]{abel}
{Abel}, T., {Bryan}, G.~L., \& {Norman}, M.~L. 2002, Science, 295, 93

\bibitem[{{Aguirre} {et~al.}(2001)}]{aguirre-full}
{Aguirre}, A., {et~al.} 2001, \apj, 561, 521

\bibitem[{{Barkana} \& {Loeb}(1999)}]{barkana99}
{Barkana}, R., \& {Loeb}, A. 1999, \apj, 523, 54

\bibitem[{{Barkana} \& {Loeb}(2001)}]{barkana01}
---. 2001, \physrep, 349, 125

\bibitem[{{Barkana} \& {Loeb}(2004)}]{barkana04}
---. 2004, \apj, 609, 474 

\bibitem[{{Becker} {et~al.}(2001)}]{becker}
{Becker}, R.~H., {et~al.} 2001, \aj, 122, 2850

\bibitem[{{Benson} {et~al.}(2004){Benson}, {Kamionkowski}, \&
  {Hassan}}]{benson04}
{Benson}, A.~J., {Kamionkowski}, M., \& {Hassan}, S.~H. 2004, \mnras, submitted
  (astro-ph/00407136)

\bibitem[{{Bond} {et~al.}(1991){Bond}, {Cole}, {Efstathiou}, \&
  {Kaiser}}]{bond91}
{Bond}, J.~R., {Cole}, S., {Efstathiou}, G., \& {Kaiser}, N. 1991, \apj, 379,
  440

\bibitem[{{Bromm} {et~al.}(2002){Bromm}, {Coppi}, \& {Larson}}]{bromm02}
{Bromm}, V., {Coppi}, P.~S., \& {Larson}, R.~B. 2002, \apj, 564, 23

\bibitem[{{Bromm} {et~al.}(2001{\natexlab{a}}){Bromm}, {Ferrara}, {Coppi}, \&
  {Larson}}]{bromm01}
{Bromm}, V., {Ferrara}, A., {Coppi}, P.~S., \& {Larson}, R.~B.
  2001{\natexlab{a}}, \mnras, 328, 969

\bibitem[{{Bromm} {et~al.}(2001{\natexlab{b}}){Bromm}, {Kudritzki}, \&
  {Loeb}}]{bromm-vms}
{Bromm}, V., {Kudritzki}, R.~P., \& {Loeb}, A. 2001{\natexlab{b}}, \apj, 552,
  464

\bibitem[{{Bromm} \& {Loeb}(2003)}]{bromm03-met}
{Bromm}, V., \& {Loeb}, A. 2003, \nat, 425, 812

\bibitem[{{Cen}(2003{\natexlab{a}})}]{cen03}
{Cen}, R. 2003{\natexlab{a}}, \apjl, 591, L5

\bibitem[{{Cen}(2003{\natexlab{b}})}]{cen03-letter}
---. 2003{\natexlab{b}}, \apj, 591, 12

\bibitem[{{Cen}(2003{\natexlab{c}})}]{cen03-shock}
---. 2003{\natexlab{c}}, \apj, submitted (astro-ph/0311329)

\bibitem[{{Ciardi} \& {Madau}(2003)}]{ciardi03}
{Ciardi}, B., \& {Madau}, P. 2003, \apj, 596, 1

\bibitem[{{Cole} {et~al.}(1994){Cole}, {Aragon-Salamanca}, {Frenk}, {Navarro},
  \& {Zepf}}]{cole94}
{Cole}, S., {Aragon-Salamanca}, A., {Frenk}, C.~S., {Navarro}, J.~F., \&
  {Zepf}, S.~E. 1994, \mnras, 271, 781

\bibitem[Cooray \& Sheth(2002)]{cooray02}
Cooray, A.~\& Sheth, R.\ 2002, \physrep, 372, 1 

\bibitem[{{Dekel} \& {Woo}(2003)}]{dekel03}
{Dekel}, A., \& {Woo}, J. 2003, \mnras, 344, 1131

\bibitem[{{Dijkstra} {et~al.}(2004a){Dijkstra}, {Haiman}, {Rees}, \&
  {Weinberg}}]{dijkstra04a}
{Dijkstra}, M., {Haiman}, Z., {Rees}, M.~J., \& {Weinberg}, D.~H. 2004, \apj,
  601, 666

\bibitem[{{Dijkstra} {et~al.}(2004b){Dijkstra}, {Haiman}, \& {Loeb}
}]{dijkstra04b}
{Dijkstra}, M., {Haiman}, Z., {Loeb}, A. 2004, \apj, 613, 646

\bibitem[{{Efstathiou}(1992)}]{efstathiou92}
{Efstathiou}, G. 1992, \mnras, 256, 43

\bibitem[Eisenstein \& Hu(1998)]{eis98}
Eisenstein, D.~J.~\& Hu, W.\ 1998, \apj, 496, 605 

\bibitem[{{Fan} {et~al.}(2002)}]{fan}
{Fan}, X., {et~al.} 2002, \aj, 123, 1247

\bibitem[{{Fukugita} \& {Kawasaki}(2003)}]{fukugita03}
{Fukugita}, M., \& {Kawasaki}, M. 2003, \mnras, 343, L25

\bibitem[{{Furlanetto} {et~al.}(2004{\natexlab{a}}){Furlanetto}, {Hernquist},
  \& {Zaldarriaga}}]{furl04c}
{Furlanetto}, S.~R., {Hernquist}, L., \& {Zaldarriaga}, M. 2004{\natexlab{a}},
  \mnras, 354, 695

\bibitem[{{Furlanetto} \& {Loeb}(2001)}]{furl-bigm}
{Furlanetto}, S.~R., \& {Loeb}, A. 2001, \apj, 556, 619

\bibitem[{{Furlanetto} \& {Loeb}(2003)}]{furl-metals}
---. 2003, \apj, 588, 18

\bibitem[{{Furlanetto} {et~al.}(2004{\natexlab{b}}){Furlanetto}, {Zaldarriaga},
  \& {Hernquist}}]{furl04a}
{Furlanetto}, S.~R., {Zaldarriaga}, M., \& {Hernquist}, L. 2004{\natexlab{b}},
  \apj, 613, 1 [FZH04]

\bibitem[{{Furlanetto} {et~al.}(2004{\natexlab{c}}){Furlanetto}, {Zaldarriaga},
  \& {Hernquist}}]{furl04b}
---. 2004{\natexlab{c}}, \apj, 613, 16

\bibitem[Gnedin \& Hui(1998)]{gnedin98}
Gnedin, N.~Y.~\& Hui, L.\ 1998, \mnras, 296, 44 

\bibitem[{{Gunn} \& {Peterson}(1965)}]{gp}
{Gunn}, J.~E., \& {Peterson}, B.~A. 1965, \apj, 142, 1633

\bibitem[{{Haiman} {et~al.}(1996){Haiman}, {Thoul}, \& {Rees}}]{haiman96} 
Haiman, Z., Thoul, A.~A., \& Loeb, A.\ 1996, \apj,
464, 523

\bibitem[{{Haiman} {et~al.}(2001){Haiman}, {Abel}, \& {Madau}}]{haiman01-mh}
{Haiman}, Z., {Abel}, T., \& {Madau}, P. 2001, \apj, 551, 599

\bibitem[{{Haiman} {et~al.}(2000){Haiman}, {Abel}, \& {Rees}}]{haiman00}
{Haiman}, Z., {Abel}, T., \& {Rees}, M.~J. 2000, \apj, 534, 11

\bibitem[{{Haiman} \& {Holder}(2003)}]{haiman03}
{Haiman}, Z., \& {Holder}, G.~P. 2003, \apj, 595, 1

\bibitem[{{Haiman} {et~al.}(1997){Haiman}, {Rees}, \& {Loeb}}]{haiman97}
{Haiman}, Z., {Rees}, M.~J., \& {Loeb}, A. 1997, \apj, 476, 458; 484, 985

\bibitem[{{Heger} \& {Woosley}(2002)}]{heger02}
{Heger}, A., \& {Woosley}, S.~E. 2002, \apj, 567, 532

\bibitem[{{Hui} \& {Haiman}(2003)}]{hui03}
{Hui}, L., \& {Haiman}, Z. 2003, \apj, 596, 9

\bibitem[{{Kauffmann} \& {White}(1993)}]{kauffmann93}
{Kauffmann}, G., \& {White}, S.~D.~M. 1993, \mnras, 261, 921

\bibitem[{{Kauffmann} {et~al.}(2003)}]{kauffmann03}
{Kauffmann}, G., {et~al.} 2003, \mnras, 341, 54

\bibitem[{{Kitayama} \& {Ikeuchi}(2000)}]{kitayama00}
{Kitayama}, T., \& {Ikeuchi}, S. 2000, \apj, 529, 615

\bibitem[{{Kogut} {et~al.}(2003)}]{kogut03}
{Kogut}, A., {et~al.} 2003, \apjs, 148, 161

\bibitem[{{Lacey} \& {Cole}(1993)}]{lacey}
{Lacey}, C., \& {Cole}, S. 1993, \mnras, 262, 627

\bibitem[{{Machacek} {et~al.}(2001){Machacek}, {Bryan}, \& {Abel}}]{machacek01}
{Machacek}, M.~E., {Bryan}, G.~L., \& {Abel}, T. 2001, \apj, 548, 509

\bibitem[{{Madau} {et~al.}(2001){Madau}, {Ferrara}, \& {Rees}}]{madau01}
{Madau}, P., {Ferrara}, A., \& {Rees}, M.~J. 2001, \apj, 555, 92

\bibitem[{{Madau} {et~al.}(1997){Madau}, {Meiksin}, \& {Rees}}]{mmr}
{Madau}, P., {Meiksin}, A., \& {Rees}, M.~J. 1997, \apj, 475, 429

\bibitem[{{Mesinger} \& {Haiman}(2004)}]{mesinger04}
{Mesinger}, A., \& {Haiman}, Z. 2004, \apjl, 611, L69

\bibitem[{{Miralda-Escud{\' e}} {et~al.}(2000){Miralda-Escud{\' e}},
  {Haehnelt}, \& {Rees}}]{miralda00}
{Miralda-Escud{\' e}}, J., {Haehnelt}, M., \& {Rees}, M.~J. 2000, \apj, 530, 1

\bibitem[{{Mo} \& {White}(1996)}]{mo96}
{Mo}, H.~J., \& {White}, S.~D.~M. 1996, \mnras, 282, 347

\bibitem[{{Mori} {et~al.}(2002){Mori}, {Ferrara}, \& {Madau}}]{mori02}
{Mori}, M., {Ferrara}, A., \& {Madau}, P. 2002, \apj, 571, 40

\bibitem[{{Oh} \& {Haiman}(2003)}]{oh03-entropy}
{Oh}, S.~P., \& {Haiman}, Z. 2003, \mnras, 346, 456

\bibitem[{{Onken} \& {Miralda-Escud{\' e}}(2004)}]{onken04}
{Onken}, C.~A., \& {Miralda-Escud{\' e}}, J. 2004, \apj, 610, 1

\bibitem[{{Press} \& {Schechter}(1974)}]{press}
{Press}, W.~H., \& {Schechter}, P. 1974, \apj, 187, 425

\bibitem[{{Rees}(1986)}]{rees86}
{Rees}, M.~J. 1986, \mnras, 222, 27P

\bibitem[{{Ricotti} {et~al.}(2002){Ricotti}, {Gnedin}, \& {Shull}}]{ricotti}
{Ricotti}, M., {Gnedin}, N.~Y., \& {Shull}, J.~M. 2002, \apj, 575, 49

\bibitem[{{Ricotti} {et~al.}(2004){Ricotti}, {Ostriker}, \&
{Gnedin}}]{ricotti04} {Ricotti}, M., {Gnedin}, N.~Y., \& {Ostriker},
J.~P. 2004, \mnras, submitted (astro-ph/0404318)

\bibitem[{{Sasaki}(1994)}]{sasaki94}
{Sasaki}, S. 1994, \pasj, 46, 427

\bibitem[{{Scalo}(1998)}]{scalo}
{Scalo}, J. 1998, in ASP Conf. Ser. 142: The Stellar Initial Mass Function, ed.
  G. Gilmore and D. Howell (San Francisco: ASP), 201

\bibitem[Scannapieco et~al.(2001)]{scann01} 
Scannapieco, E., Thacker, R.~J., \& Davis, M.\ 2001, \apj, 557, 605 

\bibitem[Scannapieco \& Barkana(2002)]{scann-bark02}
Scannapieco, E.~\& Barkana, R.\ 2002, \apj, 571, 585 

\bibitem[{{Scannapieco} {et~al.}(2002){Scannapieco}, {Ferrara}, \&
  {Madau}}]{scann02}
{Scannapieco}, E., {Ferrara}, A., \& {Madau}, P. 2002, \apj, 574, 590

\bibitem[{{Scannapieco} {et~al.}(2003){Scannapieco}, {Schneider}, \&
  {Ferrara}}]{scann03}
{Scannapieco}, E., {Schneider}, R., \& {Ferrara}, A. 2003, \apj, 589, 35

\bibitem[{{Sedov}(1959)}]{sedov}
{Sedov}, L.~I. 1959, {Similarity and Dimensional Methods in Mechanics} (New
  York: Academic Press)

\bibitem[{{Shapiro} \& {Giroux}(1987)}]{shapiro87}
{Shapiro}, P.~R., \& {Giroux}, M.~L. 1987, \apjl, 321, L107

\bibitem[{{Shapiro} {et~al.}(2004){Shapiro}, {Iliev}, \& {Raga}}]{shapiro04}
{Shapiro}, P.~R., {Iliev}, I.~T., \& {Raga}, A.~C. 2004, \mnras, 348, 753

\bibitem[{{Sheth}(1998)}]{sheth98}
{Sheth}, R.~K. 1998, \mnras, 300, 1057

\bibitem[Sokasian et al.(2002)]{sok02}
Sokasian, A., Abel, T., \& Hernquist, L.\ 2002, \mnras, 332, 601 

\bibitem[{{Sokasian} {et~al.}(2004)}]{sokasian04}
{Sokasian}, A., {et~al.} 2004, \mnras, 350, 47

\bibitem[{{Somerville} \& {Kolatt}(1999)}]{somerville99}
{Somerville}, R.~S., \& {Kolatt}, T.~S. 1999, \mnras, 305, 1

\bibitem[{{Somerville} \& {Livio}(2003)}]{somerville03}
{Somerville}, R.~S., \& {Livio}, M. 2003, \apj, 593, 611

\bibitem[{{Songaila}(2004)}]{songaila04}
{Songaila}, A. 2004, \aj, 127, 2598

\bibitem[{{Spergel} {et~al.}(2003)}]{spergel03}
{Spergel}, D.~N., {et~al.} 2003, \apjs, 148, 175

\bibitem[{{Strickland} {et~al.}(2000){Strickland}, {Heckman}, {Weaver}, \&
  {Dahlem}}]{strickland00}
{Strickland}, D.~K., {Heckman}, T.~M., {Weaver}, K.~A., \& {Dahlem}, M. 2000,
  \aj, 120, 2965

\bibitem[{{Tegmark} {et~al.}(1997)}]{tegmark97}
{Tegmark}, M., {et~al.} 1997, \apj, 474, 1

\bibitem[{{Theuns} {et~al.}(2002)}]{theuns02-reion}
{Theuns}, T., {et~al.} 2002, \apjl, 567, L103

\bibitem[{{Thoul} \& {Weinberg}(1996)}]{thoul96}
{Thoul}, A.~A., \& {Weinberg}, D.~H. 1996, \apj, 465, 608

\bibitem[{{Verde} {et~al.}(2001){Verde}, {Kamionkowski}, {Mohr}, \&
  {Benson}}]{verde01}
{Verde}, L., {Kamionkowski}, M., {Mohr}, J.~J., \& {Benson}, A.~J. 2001,
  \mnras, 321, L7

\bibitem[{{Whalen} {et~al.}(2004){Whalen}, {Abel}, \& {Norman}}]{whalen04}
{Whalen}, D., {Abel}, T., \& {Norman}, M.~L. 2004, \apj, 610, 14

\bibitem[{{White} {et~al.}(2003){White}, {Becker}, {Fan}, \&
  {Strauss}}]{white03}
{White}, R.~L., {Becker}, R.~H., {Fan}, X., \& {Strauss}, M.~A. 2003, \aj, 126,
  1

\bibitem[{{Wyithe} \& {Loeb}(2002)}]{wyithe-lumfcn}
{Wyithe}, J.~S.~B., \& {Loeb}, A. 2002, \apj, 581, 886

\bibitem[{{Wyithe} \& {Loeb}(2003{\natexlab{a}})}]{wyithe03}
---. 2003{\natexlab{a}}, \apj, 586, 693

\bibitem[{{Wyithe} \& {Loeb}(2003{\natexlab{b}})}]{wyithe03-letter}
---. 2003{\natexlab{b}}, \apjl, 588, L69

\bibitem[Wyithe \& Loeb(2003{\natexlab{c}})]{wyithe03c}
---. 2003{\natexlab{c}}, \apj, 595, 614 

\bibitem[{{Wyithe} \& {Loeb}(2004{\natexlab{a}})}]{wyithe04-prox}
---. 2004{\natexlab{a}}, \nat, 427, 815

\bibitem[{{Wyithe} \& {Loeb}(2004{\natexlab{b}})}]{wyithe04-bub}
---. 2004{\natexlab{b}}, \apj, submitted (astro-ph/0407162)

\bibitem[{{Wyithe} \& {Loeb}(2004{\natexlab{c}})}]{wyithe04-siz}
---. 2004{\natexlab{c}}, Nature, 432, 194

\bibitem[{{Zaldarriaga} {et~al.}(2004){Zaldarriaga}, {Furlanetto}, \&
  {Hernquist}}]{zald04}
{Zaldarriaga}, M., {Furlanetto}, S.~R., \& {Hernquist}, L. 2004, \apj, 608, 622

\end{thebibliography}

\end{document}